\documentclass[aps,twocolumn,prl,showpacs]{revtex4}
\usepackage{epsfig,bm,amsmath,amssymb,natbib}
\usepackage{hyperref}
\usepackage{braket}
\usepackage{color}

\newcommand{\JXf}{J_{\mbox{\tiny\it Xf}}}

\newcommand{\BSAT}{B_{\mbox{\tiny SAT}}}

\newcommand{\muB}{\mu_{\mbox{\tiny\it B}}}

\newcommand{\RPol}{R_{\mbox{\tiny Pol}}}

\newcommand{\BXf}{B_{\mbox{\tiny Xf}}}
\newcommand{\kB}{k_{\mbox{\tiny B}}}

\newcommand{\muEu}{\mu_{\mbox{\tiny Eu}}}

\newcommand{\dPol}{d_{\mbox{\tiny Pol}}}

\newcommand{\nPol}{n_{\mbox{\tiny Pol}}}
\newcommand{\DeltaThetaFSAT}{\Delta\theta_F^{\mbox{\tiny SAT}}}
\newcommand{\DeltaThetaF}{\Delta\theta_F}

\newcommand{\muPol}{\mu_{\mbox{\tiny Pol}}}
\newcommand{\muPolMFA}{\mu_{\mbox{\tiny Pol}}^{\mbox{\tiny MFA}}}

\newcommand{\tPol}{\tau_{\mbox{\tiny\rm Pol}}}

\usepackage{draftwatermark}
\SetWatermarkText{DRAFT}
\SetWatermarkScale{5}
\SetWatermarkColor[gray]{0.98}

\begin{document}
\title{Ultrafast light switching of ferromagnetism in EuSe}
\author{A. B. Henriques,$^{1}$ X. Gratens,$^{1}$
P. A. Usachev,$^{1}$\footnote{Permanent address: Ioffe Institute, 194021 St. Petersburg, Russia}
V. A.  Chitta,$^{1}$ and G. Springholz$^2$}
\address{$^1$ Instituto de F\'{\i}sica, Universidade de S\~ao Paulo, 05508-090 S\~ao Paulo, Brazil\\
$^2$Institut f\"ur Halbleiter und Festk\"orperphysik, Johannes Kepler Universit\"at Linz, 4040 Linz, Austria}
\begin{widetext}
\begin{abstract}
We demonstrate that light resonant with the bandgap forces the antiferromagnetic semiconducor EuSe to enter ferromagnetic alignment in the picosecond time scale.
A photon generates an electron-hole pair,
whose electron forms a supergiant spin polaron of magnetic moment of nearly 6,000 Bohr magnetons.
By increasing the light intensity,
the whole of the sample can be fully magnetized.
The key to the novel large photoinduced magnetization mechanism is the huge enhancement of the magnetic susceptibility
when both antiferromagnetic and ferromagnetic interactions are present in the material, and are of nearly equal magnitude, as is the case in EuSe.
\end{abstract}
\date{\today}
\end{widetext}
\maketitle

The  ultrafast control of the magnetic state of matter is a topic of vast current interest for the development of applications and for advancing the knowledge in the field of light-matter interaction \cite{SciRepLightControlledZnO,NatComm2017,Nature2011}.
Here we report on the discovery of an ultrafast mechanism of light switching of
antiferromagnetic EuSe into the ferromagnetic phase, which is efficient in the proximity of its N\'eel temperature.

\begin{figure}
\includegraphics[angle=0,width=60mm]{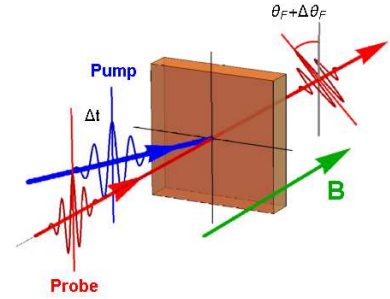}
\caption{
Scheme of the setup for measuring the time-resolved photoinduced Faraday rotation angle. The linearly polarized probe pulse arrives at the sample when a time $\Delta t$ has elapsed after the arrival of the pump pulse. We measured $\DeltaThetaF$, the Faraday rotation of the probe, induced by the pump illumination.
}
\label{fig:setup}
\end{figure}

The EuSe samples were grown by molecular beam epitaxy
(MBE) onto (111) BaF$_2$ substrates. Because of the almost
perfect lattice constant matching ($a=6.191~{\mbox{\AA}}$ and $a=6.196~{\mbox{\AA}}$ for EuSe and BaF$_2$, respectively), nearly
unstrained bulklike EuSe reference layers with $\mu$m thickness were obtained directly by growth on BaF$_2$.
The time-resolved photoinduced Faraday was measured using a two-color pump-probe technique using 2~ps light pulses, as illustrated in Fig.~\ref{fig:setup}.
The Faraday rotation angle of the probe light pulse, $\DeltaThetaF$, induced by the pump pulse, was measured using lock-in techniques with a resolution better than 10$^{-7}$~radians. The image of the excitation spot on the sample had a diameter of 150~$\mu$m, about twice the diameter of the probe spot. For continuous wave (CW) measurements the same setup was used, except that the pump source was a doubled Nd:Yag laser or a Xe lamp coupled to a monochromator, and the probe source was a semiconductor laser of energy below the EuSe bandgap. All measurements were performed using an optical cryostat containing a superconducting coil for magnetic fields up to 8~Teslas, applied in the Faraday geometry.
\begin{figure}
\includegraphics[angle=0,width=90mm]{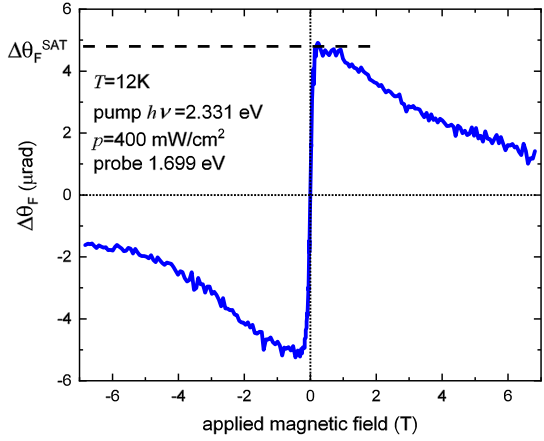}
\caption{
Photoinduced Faraday rotation in EuSe. The maximum PFR is indicated by $\DeltaThetaFSAT$.
}
\label{fig:PFRHighB}
\end{figure}
\begin{figure}
\includegraphics[angle=0,width=90mm]{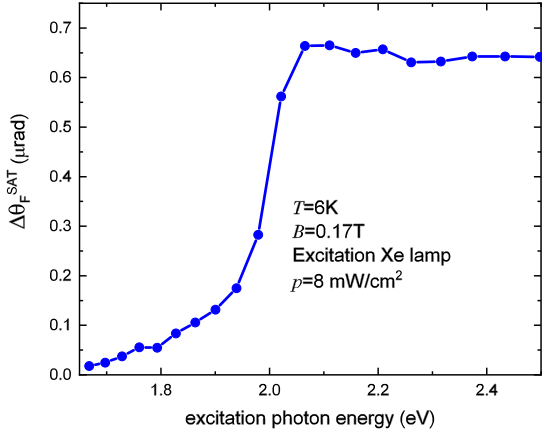}
\caption{
PFR excitation spectrum.
}
\label{fig:ExcitationPFR}
\end{figure}
Figure \ref{fig:PFRHighB} shows the photoinduced Faraday rotation (PFR) as a function of applied magnetic field, for CW excitation with intensity $p=400$~mW/cm$^{-2}$, at $T=12$~K.
The PFR signal has all the characteristics expected for photoinduced magnetic polarons: no signal at zero magnetic field, because polarons are radomly oriented and produce zero magnetization, and a very rapid increase when a magnetic field is applied,
due to the complete alignment of the polaron ensemble along the field due to the Zeeman torque acting on particles with a large magnetic moment. Fig.~\ref{fig:PFRHighB} shows that when the applied magnetic field is increased beyond 1T, the
PFR effect gradually decreases towards zero; this is because large magnetic fields progressively force the EuSe lattice spins into ferromagnetic order, and ultimately the formation of spin polarons is quenched.
Fig.~\ref{fig:ExcitationPFR} shows that the PFR signal vanishes if the excitation photon energy is less than the bandgap, and increases abruptly at the bandgap energy, demonstrating that the PFR effect is associated with photogenerated conduction band electrons.
\begin{figure}
\includegraphics[angle=0,width=90mm]{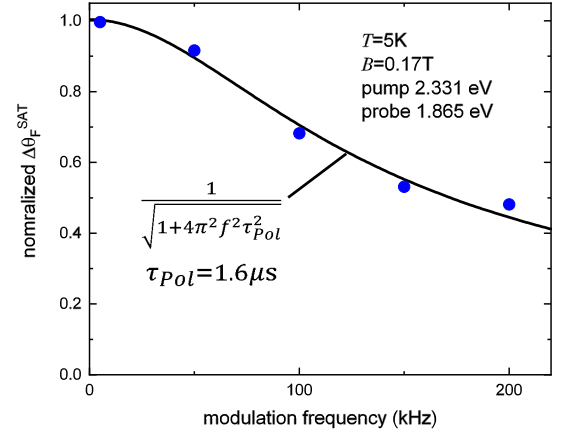}
\caption{
PFR as a function of the pump modulation frequency.
}
\label{fig:TauPFR}
\end{figure}

Because the photoexcited electrons are bound to very heavy photoexcited holes \cite{prb05}, the spin polarons are immobile and are confined to a
layer below the surface of the crystal whose thickness equals the penetration depth of the excitation light, $1/\alpha$,
where $\alpha\cong 15\,\mu$m$^{-1}$ is the EuSe absorption coefficient for the excitation photon energy  \cite{prb05}.
From elementary kinetics, the steady-state density of photoinduced polarons is given by\cite{prb17a}
\begin{equation}
\nPol=\chi\frac{p\alpha\tPol}{h\nu}\leq \frac{p\alpha\tPol}{h\nu},
\label{eq:nPol}
\end{equation}
where $\chi\leq 1$ is the quantum efficiency for spin polaron photogeneration, and $\tPol$ is the spin polaron lifetime. The spin polaron lifetime was deduced from the dependence of the PFR amplitude on the
modulation frequency of the excitation light, shown in Fig.~\ref{fig:TauPFR}, following the procedure described in \onlinecite{prb16}, giving $\tPol$=1.6~$\mu$sec. The mean distance between polarons, $d$, in units of the
EuSe lattice constant, is therefore
\begin{equation}
\frac{d}{a}\geq \frac{2}{a}\left(\frac{3}{4\pi}\frac{h\nu}{p\alpha\tPol}\right)^{1/3}
\label{eq:dPol}
\end{equation}

\begin{figure}
\includegraphics[angle=0,width=90mm]{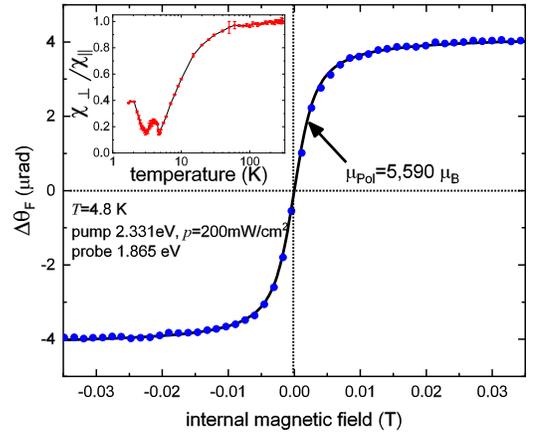}
\caption{
PFR as a function of internal magnetic field.
The ratio $\chi_\perp/\chi_{||}$, used to compute the internal magnetic field, is shown in the inset, as a function of temperature.
}
\label{fig:PFRmuPol}
\end{figure}

Substituting all parameters into \eqref{eq:dPol}, we obtain $\dPol\geq 36\,a$. Thus, $d$ is at least one order of magnitude greater than the typical radius of a spin polaron in europium chalcogenide \cite{apl2011}, $\RPol\sim 3ã$. Being so far apart, the polarons are non-interacting.
Moreover, above the N\'eel temperature the spin polarons orientation floats freely, therefore
they form a superparamagnetic gas, whose magnetization dependence on magnetic field and temperature obeys a Langevin function \cite{bean}. Because PFR is proportional to the magnetization \cite{prb17b}, the photoinduced Faraday rotation angle will be given by
\begin{equation}
\DeltaThetaF=\DeltaThetaFSAT L\left(\frac{\muPol B}{\kB T}\right)
\label{eq:Langevin}
\end{equation}

The magnitude of the magnetic moment of the spin polaron at a given temperature can be determined accurately by fitting the experimental curve with \eqref{eq:Langevin}, because it is the sole adjustable parameter determining the sharpness of the step. However, we must first convert the applied magnetic field into internal one,
which is smaller than the former due to the demagnetizing field \cite{blundell}.
For the Faraday geometry used here, where the magnetic field is normal to a very thin epitaxial layer, the ratio of the internal magnetic field to the applied one is equal to the ratio
of the magnetic susceptibility measured when the applied field is perpendicular to the layer, ($\chi_\perp$), to the susceptibility measured when the applied field is parallel to it ($\chi_{||}$).
The susceptibilities $\chi_\perp$ and $\chi_{||}$ were measured using a SQUID magnetometer, which had a magnetic moment resolution better than 10$^{-11}$~Am$^2$.
Figure \ref{fig:PFRmuPol} shows the measured PFR as a function of internal magnetic field. The inset shows the measured ratio between susceptibilities $\chi_\perp$ and $\chi_\parallel$, used to convert the applied magnetic field into internal one, above the N\'eel temperature. The solid line in Fig.~\ref{fig:PFRmuPol} shows the fit of the theory, using \eqref{eq:Langevin}, which yields the magnetic moment of the spin polaron, $\muPol=5,590$ Bohr magnetons ($\muB$). This magnetic moment is 10 times greater than the giant photoinduced spin polarons known so far \cite{prb16,prb17a}, which allows us to categorize the spin polarons discovered as supergiant.
\begin{figure}
\includegraphics[angle=0,width=90mm]{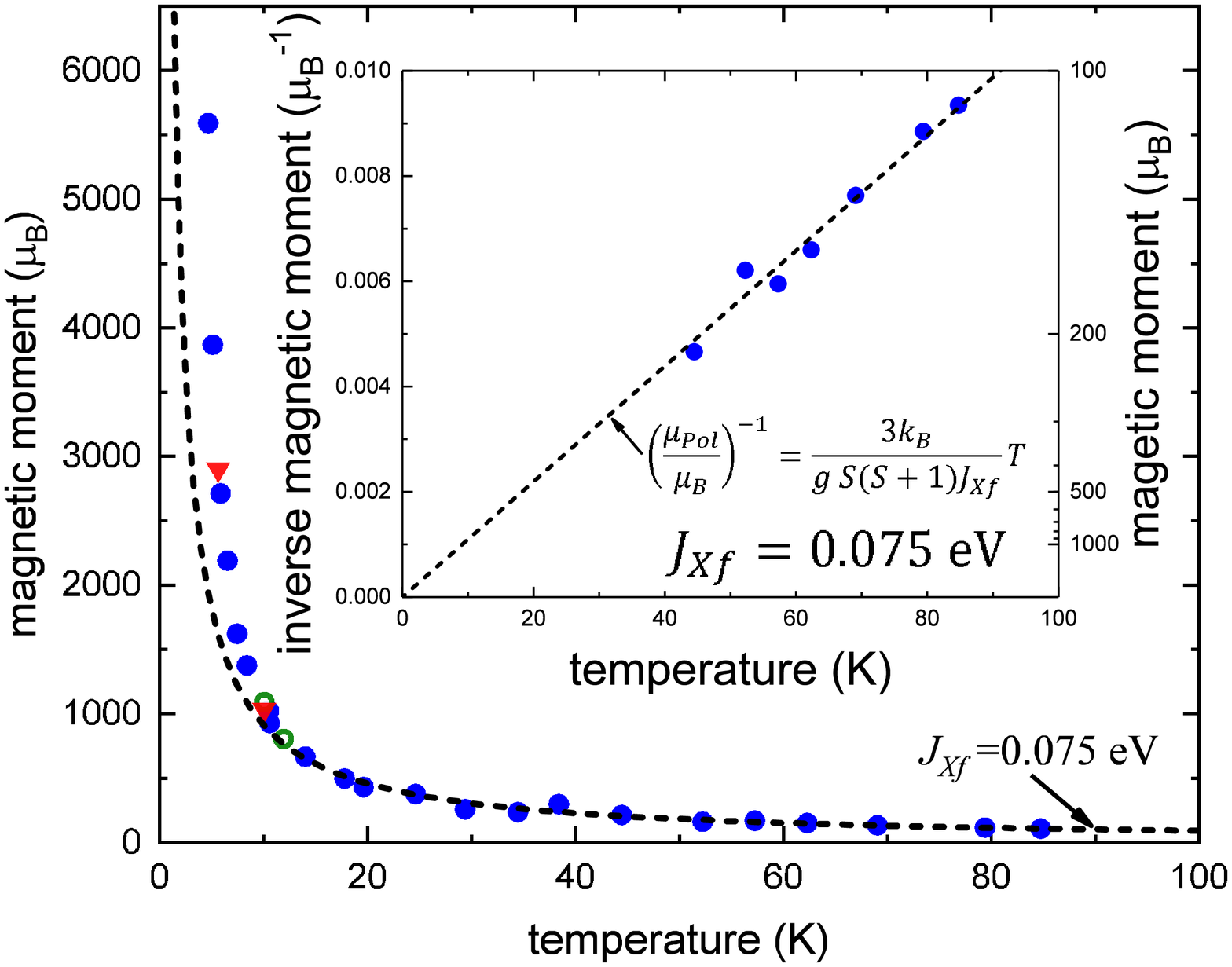}
\caption{
Magnetic moment of a photoinduced polaron as a function of temperature. Full circles, empty circles and inverted triangles correspond to probe wavelength of 665, 730 and 760~nm, respectively.
The dashed line shows the paramagnetic approximation, with $\JXf=0.075$~eV obtained from the linear fit shown in the inset.}
\label{fig:muPolVsT}
\end{figure}
The magnetic moment of the spin polaron as a function of temperature is shown in Fig.~\ref{fig:muPolVsT}. Increasing the temperature quenches spin polaron formation, due to the decreasing magnetic susceptibility of the lattice spins. At temperatures much higher than the N\'eel temperature, and $\muEu B/k_BT\ll 1$, the magnetization of lattice spins can be well described by the paramagnetic approximation
\begin{equation}
M=N\muEu\frac{S+1}{3S}\frac{\muEu B}{k_BT}.
\label{eq:paramagneticMagnetization}
\end{equation}
where $N=4/a^3$ is the density of Eu sites in the crystal and $\muEu=g\muB S$ is the magnetic moment of an Eu atom, where $g=2$, $S=7/2$.
The exchange interaction between the photoexcited electron, whose wavefunction is $\psi(r)$, and the lattice spins,
is described by an effective magnetic field, $\BXf$, acting on the lattice spins \cite{prb14}
\begin{equation}
\BXf=\frac{\JXf\,S}{N\muEu}|\psi(r)|^2,
\label{eq:BXf}
\end{equation}
where $\JXf$ is the exchange interaction integral between the electron forming the spin polaron and the lattice spins \cite{apl2011}.
Substituting \eqref{eq:BXf} in \eqref{eq:paramagneticMagnetization},
and integrating in the whole volume, the high temperature magnetic moment of a spin polaron in the paramagnetic crystal approximation is obtained
\begin{equation}
\muPol(T)=g\muB S(S+1)\frac{\JXf}{3k_BT},
\label{eq:muPolhighT}
\end{equation}
A fit of the high temperature tail, $T>40$~K, of $\muPol$ versus $T$ with \eqref{eq:muPolhighT}, whereby $\JXf$ is the only adjustable parameter, is shown in the inset of Fig.~\ref{fig:muPolVsT}, and gives $\JXf=75$~meV. This $\JXf$ value is about the same as measured for EuTe, meaning that the mean exchange field generated by the photoexcited electron in EuSe is about the same as for EuTe, i.e. about 1~Tesla \cite{apl2011}.
Fig.~\ref{fig:muPolVsT} shows that when the temperature is decreased below 20K and approaches the N\'eel temperature, $\muPol$ increases much faster than the paramagnetic
approximation, and can reach {\em almost an order of magnitude greater} than $\muPol$ that would be observed if the exchange interaction was switched off (the paramagnetic approximation).
This is in sharp contrast to EuTe, where $\muPol$ {\em is always smaller} than the paramagnetic limit \cite{prb16}. The reason for the very large
increase in EuSe is the near equal absolute values of the first and second neighbor exchange interaction constants, $J_1$ and $J_2$. Whereas $J_1>0$ favors
ferromagnetism, $J_2<0$ favors antiferromagnetism, which predominates only because $|J_2|$ is marginally larger than $|J_1|$. This also implies
than around the N\'eel temperature only a very small internal magnetic field, of about 0.1~T is sufficient to promote a phase transition into the ferromagnetic state, as demonstrated experimentally in Ref.~\onlinecite{springholz}. Bearing in mind that within a spin polaron the effective magnetic field acting on the spins is about 1~T,
this implies that the within the polaron the crystal lattice attains ferromagnetic alignment, which explains why the magnetic moment of the polaron is so
large in EuSe. Such reasoning is supported quantitatively by the Weiss mean field approximation (MFA) with first and second neighbor interaction at $T=0$K, according to which in the AFM-II phase the magnetic field to induce ferromagnetic alignment is given by \cite{prb14,springholzmcarlo,apl2011}
\begin{equation}
\BSAT=\frac{24 \left|J_1+J_2\right|S}{g\mu_B},
\label{eq:bsat}
\end{equation}
and in the range $B<\BSAT$ the magnetization is given by $M=N\muEu B/\BSAT$. Using \eqref{eq:BXf} and \eqref{eq:bsat}, and integrating $M$ in the volume, the magnetic moment of the spin polaron
in the mean field approximation is obtained
\begin{equation}
\muPolMFA\mbox{($T=0$K)}=\frac{\JXf}{24\left|J_1+J_2\right|}g\muB.
\label{eq:mupolMFA}
\end{equation}
Table \ref{tab:pars} displays the result produced by formula \eqref{eq:mupolMFA} for EuTe and EuSe.
The calculated values agree by order of magnitude with the measured values, and demonstrate that the very large $\muPol$ in EuSe is due to the near cancellation of the ferro and antiferromagnetic lattice interactions.
\setlength{\tabcolsep}{1pt}
\begin{table}[h]
\caption{EuSe and EuTe parameters, $\muPol$ measured at $T=5$~K and estimated using the MFA for $T=0$~K.}
\small
\centering
\begin{tabular}{cccccc}
\hline\hline
 & & & & \multicolumn{2}{c}{$\muPol\,(\muB)$} \\
 & $J_1$(K) & $J_2$(K) & $\JXf$(eV) & Experiment & MFA \\
\hline
 EuTe & 0.043 & -0.15 & 0.083 & 600 & 750\\
 Ref. & \multicolumn{2}{c}{\onlinecite{Wachter}} & \onlinecite{prb14} & \onlinecite{prb17a}\\
\hline
EuSe & 0.29 & -0.30 & 0.075 & 5,590 & 7,300 \\
Ref. & \multicolumn{2}{c}{\onlinecite{Xavier}} & \multicolumn{2}{c}{this work}\\
\hline\hline
\end{tabular}
\label{tab:pars}
\end{table}

The MFA does not provide any information about the internal structure of the spin polaron. A more accurate description of the spin polaron, and a description of its internal structure, is provided by a self-consistent calculation for an AFM-II type system at $T=0$~K \cite{prb14}.
In performing such a calculation for EuSe, the input parameters were a conduction band effective mass $m^\ast=0.3 m_0$ \cite{cho}, a dielectric constant $\varepsilon=9.4$ \cite{mauger}, as well as the parameters given in Table \ref{tab:pars}. The self-consistent calculation produces a spin polaron containing a large ferromagnetic core of radius 2.8 lattice parameters, containing 370 ferromagnetically aligned Eu atoms, with a calculated spin polaron of magnetic moment of $\muPol=4,920\,\mu_B$.

\begin{figure}
\includegraphics[angle=0,width=90mm]{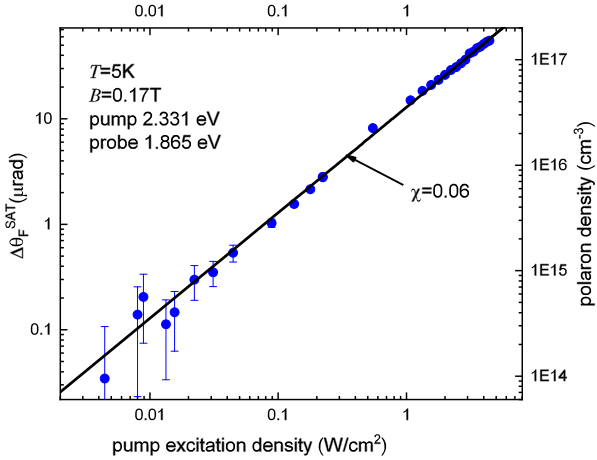}
\caption{
PFR as a function of pump excitation power.}
\label{fig:nPolVsPower}
\end{figure}
Figure \ref{fig:nPolVsPower} shows $\DeltaThetaFSAT$, and the corresponding polaron population, deduced using the Verdet constant as described in Ref.\onlinecite{prb17b},
as a function of excitation pump power density. The solid line is a fit using \eqref{eq:nPol} with $\chi$ as the only fitting parameter, which yields $\chi=0.053$. In dramatic contrast to EuTe, which shows a saturation of the polaron population, in EuSe the population grows linearly with excitation even at concentrations far above expected residual deffect concentrations, indicating that the photoinduced polarons are intrinsic, and therefore is should be possible to fully magnetize the layer penetrated by light simply by
using enough excitation power. The maximum pump intensity shown in Fig.\ref{fig:nPolVsPower} generated about $2\times 10^{17}$~cm$^{-3}$ spin polarons, and taking into account their ferromagnetic core radius of about 3~a, this corresponds to a magnetization of $5\times10^{-3}$ of the saturation value. We could not exploit greater pump intensities without heating of the sample, because we used an optical chopper with a high (50\%) duty cycle.

\begin{figure}
\includegraphics[angle=0,width=90mm]{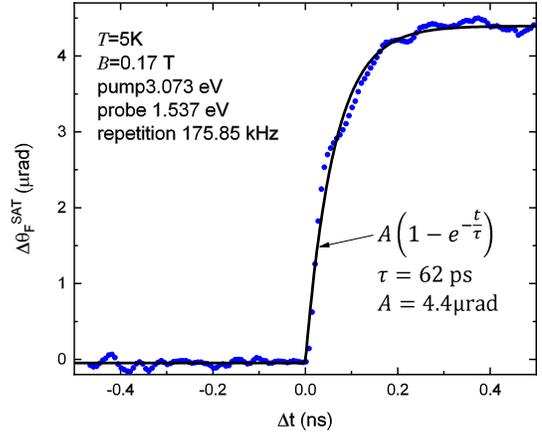}
\caption{
PFR as a function of the delay, $\Delta t$, between pump and probe pulses.
}
\label{fig:PFRvsDelay}
\end{figure}

Figure~\ref{fig:PFRvsDelay} shows the PFR signal as a function of the delay, $\Delta t$, between the pump and probe pulses (Fig.\ref{fig:setup}). Before the arrival of the pump pulse (negative delay), no PFR signal is detected, however, after
the arrival of the pump pulse the PFR signal grows exponentially with a characteristic rise time of 62~ps, which classifies the photomagnetization process as ultrafast. After achieving the maximum, the PFR remained approximately constant within the maximum delay available of our experiment, of a few nanoseconds, which is explained by the long lifetime of the polarons, $\tPol=1.6\,\mu$s, reported above.

In conclusion, we have demonstrated that light can be used to convert a zero magnetization state of a EuSe crystal into a completely polarized ferromagnetic state in the ultrafast time scale,
through the photogeneration of supergiant intrinsic spin polarons. This magetization mechanism is made possible because the exchange
interaction between lattice spins in EuSe contains both ferromagnetic and antiferromagnetic components, which nearly cancel each other out.

This work was supported by the Brazilian agencies CNPq
(Project 304685/2010-0) and FAPESP (Project  2016/24125-5).

\end{document}